\documentclass[12pt,journal,final,onecolumn]{IEEEtran}

\usepackage[dvips]{graphicx}
\usepackage{epsfig}
\usepackage{amsmath}
\usepackage{amssymb}
\usepackage{amsthm}
\usepackage{amsfonts}
\usepackage{bm}
\usepackage{dsfont}
\usepackage{color}
\usepackage[mathscr]{eucal}
\usepackage{cite}

\graphicspath{{figs/}}

\newcommand{\tsf}[1]{\textsf{#1}}

\newcommand{\defeq}{\triangleq}
\newcommand{\Pp}{\mathds{P}}
\newcommand{\E}{\mathds{E}}
\newcommand{\N}{\mathds{N}}

\newcommand{\R}{\mathds{R}}
\newcommand{\Rp}{\R_{+}}

\newcommand{\abs}[1]{\lvert{#1}\rvert}
\newcommand{\card}[1]{\abs{#1}}

\newcommand{\lambdauc}{\lambda^{\textup{\tsf{UC}}}}
\newcommand{\Lambdauc}{\Lambda^{\textup{\tsf{UC}}}}
\newcommand{\lambdamc}{\lambda^{\textup{\tsf{MC}}}}
\newcommand{\Lambdamc}{\Lambda^{\textup{\tsf{MC}}}}

\newcommand{\tlambdauc}{\tilde{\lambda}^{\textup{\tsf{UC}}}}

\newcommand{\tlambdamc}{\tilde{\lambda}^{\textup{\tsf{MC}}}}

\newcommand{\hLambdauc}{\hat{\Lambda}^{\textup{\tsf{UC}}}}

\newcommand{\hLambdamc}{\hat{\Lambda}^{\textup{\tsf{MC}}}}

\newtheorem{lemma}{Lemma}

\newtheorem{theorem}[lemma]{Theorem}

\theoremstyle{definition}
\newtheorem{egdummy}{Example}
\newenvironment{example}{%
    \begin{egdummy}%
        \upshape%
        }
{\qed%
\end{egdummy}}

\theoremstyle{remark}

\begin{document}

\bibliographystyle{ieeetr}

\title{Interference Alignment in Dense Wireless Networks} 

\author{Urs~Niesen%
\thanks{U. Niesen is with the Mathematics of Networks and Communications
Research Department, Bell Labs, Alcatel-Lucent. 
Email: urs.niesen@alcatel-lucent.com}%
}

\maketitle

\begin{abstract}
    We consider arbitrary dense wireless networks, in which $n$ nodes
    are placed in an arbitrary (deterministic) manner on a square region
    of unit area and communicate with each other over Gaussian fading
    channels. We provide inner and outer bounds for the $n\times
    n$-dimensional unicast and the $n\times 2^n$-dimensional multicast
    capacity regions of such a wireless network. These inner and outer
    bounds differ only by a factor $O(\log(n))$, yielding a fairly tight
    scaling characterization of the entire regions. The communication
    schemes achieving the inner bounds use interference alignment as a
    central technique and are, at least conceptually, surprisingly
    simple.
\end{abstract}


\section{Introduction}
\label{sec:intro}

Interference alignment is a recently introduced technique to cope with
the transmissions of interfering users in wireless systems see
\cite{cadambe08,maddah-ali08,nazer09}. In this paper, we apply this
technique to obtain fairly precise (up to $O(\log(n))$ factor)
information-theoretic scaling results for the unicast and multicast
capacity regions of dense wireless networks.

\subsection{Related Work}
\label{sec:intro_prior}

The study of scaling laws for wireless networks, describing the system
performance in the limit of large number of users, was initiated by
Gupta and Kumar in \cite{gupta00a}. They analyzed a network scenario in
which $n$ nodes are placed uniformly at random on a square of area one (called
a \emph{dense network} in the following) and are randomly paired into $n$
source-destination pairs with uniform rate requirement. Under a
so-called \emph{protocol channel model}, in which only point-to-point
communication is allowed and interference is treated as noise, they
showed that the largest uniformly achievable per-node rate scales as
$\Theta(n^{-1/2})$ up to a polylogarithmic factor in $n$.
Achievability was shown using a multi-hop communication scheme combined
with straight-line routing. Different constructions achieving slightly
better scaling laws, i.e., improving the polylogarithmic factor in $n$,
were subsequently presented in \cite{kulkarni04, franceschetti07a}.

These results are in some sense negative, in that they show that with
current technology, captured by the protocol channel model assumption,
the per-node rate in large wireless networks decreases with increasing
network size even if the deployment area is kept constant. An immediate
question is therefore if this negative result is due to the protocol
channel model assumption or if there is a more fundamental reason for
it. To address this question, several authors have considered an
information-theoretic approach to the problem, in which the channel is
simply assumed to be a Gaussian fading channel without any restrictions
on the communication scheme
\cite{gupta03,kramer05,xie05,aeron07,ozgur07b}. We shall refer to this
as the \emph{Gaussian fading channel model} in the following. These works
construct cooperative communication schemes and show that they can
significantly outperform multi-hop communication in dense networks. In
particular, {\"O}zg{\"u}r et al. showed in \cite{ozgur07b} that in
Gaussian fading dense wireless networks with randomly deployed nodes and
random source-destination pairing, the maximal uniformly achievable
per-node rate scales like\footnote{The notation
$\Theta(n^{\pm\varepsilon})$ is used to indicate that the maximal
uniformly achievable per-node rate is upper bounded by
$O(n^{\varepsilon})$ and lower bounded by $\Omega(n^{-\varepsilon})$.
Similar expressions will be used throughout this section.}
$\Theta(n^{\pm\varepsilon})$ for any $\varepsilon>0$.  In other words,
in dense networks\footnote{We point out that the situation is quite
different in \emph{extended networks}, in which $n$ nodes are placed on
a square of area $n$. Here network performance depends on the
\emph{path-loss exponent} $\alpha$, governing the speed of decay of
signal power as a function of distance. For small $\alpha$, cooperative
communication is order optimal, whereas for large $\alpha$, multi-hop
communication is order optimal
\cite{xie04,jovicic04,xue05,leveque05,ahmad06,xie06,ozgur07a,ozgur07b,franceschetti09}.},
cooperative communication can increase achievable rates to almost
constant scaling in $n$---significantly improving the $\Theta(n^{-1/2})$
scaling resulting from the protocol channel model assumption. The
$\Theta(n^{\pm\varepsilon})$ scaling law was subsequently tightened to
$n^{\pm\Theta(\log^{-1/2}(n))}$ in \cite{niesen09,ghaderi09}.

While these results removed the protocol channel model assumption made
in \cite{gupta00a}, they kept the assumptions of \emph{random} node
placement and \emph{random} source-destination pairing with uniform
rate. Wireless networks with \emph{random} node placement and
\emph{arbitrary} traffic pattern have been analyzed in
\cite{subramanian07, subramanian08} for the protocol channel model and
in \cite{niesen08} for the Gaussian fading channel model. On the other
hand, wireless networks with \emph{arbitrary} node placement and
\emph{random} source-destination pairing with uniform rate have been
investigated in \cite{madan08} for the protocol channel model and in
\cite{niesen09} for the Gaussian fading channel model. While methods
similar to the ones developed in \cite{madan08} can also be used to
analyze wireless networks with \emph{arbitrary} node placement and
\emph{arbitrary} traffic pattern under the protocol channel model, the
performance of such general networks under a Gaussian channel model
(i.e., an information-theoretic characterization of achievable rates) is
unknown.

Finally, it is worth mentioning \cite{jafar09,johnson09}, which derive
scaling laws for large dense interference networks. In particular,
\cite{johnson09} considers a dense random node placement with random
source-destination pairing. However, the model there is an interference
channel as opposed to a wireless network as modeled in the works
mentioned above. In other words, the source nodes cannot communicate
with each other, and similarly the destination nodes cannot communicate
with each other. This differs from the model adopted in this paper and
the works surveyed so far, in which no such restrictions are imposed.
For such interference networks, \cite{johnson09} derives the asymptotic
sum-rate as the number of nodes in the network increases.

\subsection{Summary of Results}
\label{sec:intro_contributions}

In this paper, we consider the general problem of determining achievable
rates in dense wireless networks with arbitrary node placement and
arbitrary traffic pattern. We assume a Gaussian fading channel model,
i.e., the analysis is information-theoretic, imposing no restrictions on
the nature of communication schemes used. We analyze the $n\times
n$-dimensional unicast capacity region $\Lambdauc(n)\subset\Rp^{n\times
n}$, and the $n\times 2^n$-dimensional multicast capacity region
$\Lambdamc(n)\subset\Rp^{n\times 2^n}$ of an arbitrary dense wireless
network. $\Lambdauc(n)$ describes the collection of all achievable
unicast traffic patterns (in which each message is to be sent to only
one destination node), while $\Lambdamc(n)$ describes the collection of
all achievable multicast traffic patterns (in which each message is to
be sent to a set of destination nodes). We provide explicit
approximations $\hLambdauc(n)$ and $\hLambdamc(n)$ of $\Lambdauc(n)$ and
$\Lambdamc(n)$ in the sense that
\begin{align*}
    \hLambdauc(n) & \subset \Lambdauc(n) \subset K_1\log(n)\hLambdauc(n), \\
    \hLambdamc(n) & \subset \Lambdamc(n) \subset K_2\log(n)\hLambdamc(n),
\end{align*}
for constants $K_1,K_2$ not depending on $n$. In other words,
$\hLambdauc(n)$ and $\hLambdamc(n)$ approximate the unicast and
multicast capacity regions $\Lambdauc(n)$ and $\Lambdamc(n)$ up to a
factor $O(\log(n))$. This provides tight scaling results for arbitrary
traffic pattern and arbitrary node placement.

The results presented in this paper improve the known results in several
respects. First, as already pointed out, they require no probabilistic
modeling of the node placement or traffic pattern, but rather are valid
for any node placement and any traffic pattern and include the results for
random node placement and random source-destination pairing with uniform
rate as a special case. Second, they provide information-theoretic
scaling results that are considerably tighter than the best previously
known, namely up to a factor $O(\log(n))$ here as compared to
$O(n^{\varepsilon})$ in \cite{ozgur07b} and $n^{O(\log^{-1/2}(n))}$ in
\cite{niesen09,ghaderi09}. Moreover, the results in this paper provide
an explicit expression for the pre-constant in the $O(\log(n))$ term
that is quite small, and hence these  bounds yield good results also for
small and moderate sized wireless networks. Third, the achievable scheme
used to prove the inner bound in this paper is, at least conceptually,
quite simple, in that the only cooperation needed between users is to
perform interference alignment. This contrasts with the communication
schemes achieving near linear scaling presented so far in the
literature, which require hierarchical cooperation and are harder to
analyze.

\subsection{Organization}
\label{sec_intro_organization}

The remainder of this paper is organized as follows. Section~\ref{sec:model}
introduces the network model and notation. Section~\ref{sec:main} presents the
main results of this paper. Section~\ref{sec:schemes} describes the
communication schemes used to prove achievability. Section~\ref{sec:proofs}
contains proofs, and Sections \ref{sec:discussion} and \ref{sec:conclusions}
contain discussions and concluding remarks.

\section{Network Model and Notation}
\label{sec:model}

Let 
\begin{equation*}
    A \defeq [0,1]^2
\end{equation*}
be a square of area one, and consider $n$ nodes $V(n)\subset A$ (with
$\card{V(n)}=n$) placed in an arbitrary manner on $A$. Let
$r_{u,v}$ be the Euclidean distance between nodes $u$ and $v$, and
define 
\begin{equation*}
    r_{\min}(n) \defeq n^{1/2}\min_{u\neq v} r_{u,v}.
\end{equation*}
The minimum separation between nodes in the node placement $V(n)$ is
then $r_{\min}(n)n^{-1/2}$. Note that $r_{\min}(n) = 1$ for a grid
graph, and $r_{\min}(n) \geq n^{-1}$ with high probability for $n$ nodes
placed uniformly and independently at random on $A$. In general, we have
\begin{equation}
    \label{eq:rmin}
    r_{\min}(n) \leq 4/\sqrt{\pi} < 3,
\end{equation}
and, while the results presented in this paper hold for any
$r_{\min}(n)$, the case of interest is when $r_{\min}(n)$ decays at most
polynomially with $n$, i.e., $r_{\min}(n) \geq n^{-\kappa}$ for some
constant $\kappa \geq 0$.  Note that we do not make any probabilistic
assumptions on the node placement, but rather allow an arbitrary
(deterministic) placement of nodes on $A$.  In particular, the arbitrary
node placement model adopted here contains the random node placement
model as a special case. The arbitrary node placement model is, however,
considerably more general since it allows for classes of node
placements that only appear with vanishing probability under random node
placement (e.g., node placements with large gaps or isolated nodes).

We assume the following complex baseband-equivalent channel model. The
received signal $y_v[t]$ at node $v$ at time $t$ is given by
\begin{equation*}
    y_v[t] \defeq \sum_{u\neq v}h_{u,v}[t]x_u[t]+z_v[t],
\end{equation*}
where $h_{u,v}[t]$ is the channel gain from node $u$ to node $v$,
$x_u[t]$ is the signal sent by node $u$, and $z_v[t]$ is additive
receiver noise at node $v$, all at time $t$. The additive noise
components $\{z_v[t]\}_{v,t}$ are assumed to be independent and
identically distributed (i.i.d.) circularly-symmetric complex Gaussian
random variables with mean zero and variance one. The channel gain
$h_{u,v}[t]$ has the form
\begin{equation}
    \label{eq:phase}
    h_{u,v}[t] 
    \defeq r_{u,v}^{-\alpha/2}\exp(\sqrt{-1}\theta_{u,v}[t]),
\end{equation}
where $\alpha\geq 2$ is the \emph{path-loss exponent}. As a function of
$u$ and $v$, the phase shifts $\{\theta_{u,v}[t]\}_{u,v}$ are assumed to
be i.i.d. uniformly distributed over $[0,2\pi)$. As a function of time
$t$, we only assume that $\{\theta_{u,v}[t]\}_t$ varies in a stationary
ergodic manner as a function of $t$ for every $u,v\in V(n)$. Note that
the distances $r_{u,v}$ between the nodes do not change as a function of
time and are assumed to be known throughout the network. The phase
shifts $\{\theta_{u,v}[t]\}_{u,v}$ are assumed to be known at time $t$
at every node in the network. Together with the knowledge of the
distances $\{r_{u,v}\}_{u,v}$, this implies that full causal channel
state information (CSI) is available throughout the network. We impose a
unit average power constraint on the transmitted signal $\{x_u[t]\}_t$
at every node $u$ in the network.

The phase-fading model~\eqref{eq:phase} is adopted here for consistency
with the capacity-scaling literature. All results presented in this
paper can be extended to Rayleigh fading, see
Section~\ref{sec:discussion_rayleigh}.

A \emph{unicast traffic matrix} $\lambdauc\in\Rp^{n\times n}$ associates
with every node pair $(u,w)\in V(n)\times V(n)$ the rate
$\lambdauc_{u,w}$ at which node $u$ wants to transmit a message to node
$w$. The messages corresponding to distinct $(u,w)$ pairs are assumed to
be independent.  Note that we allow the same node $u$ to be source for
several destinations $w$, and the same node $w$ to be destination for
several sources $u$.  The \emph{unicast capacity region}
$\Lambdauc(n)\subset\Rp^{n\times n}$ is the closure of the collection of
all achievable unicast traffic matrices $\lambdauc\in\Rp^{n\times n}$.
Knowledge of the unicast capacity region $\Lambdauc(n)$ provides hence
information about the achievability of any unicast traffic matrix
$\lambdauc$.

A \emph{multicast traffic matrix} $\lambdamc\in\Rp^{n\times 2^n}$
associates with every pair of node $u\in V(n)$ and subset $W\subset
V(n)$ the rate $\lambdamc_{u,W}$ at which node $u$ wants to multicast a
message to the nodes in $W$, i.e., every node $w\in W$ wants to receive
the same message from $u$. The messages corresponding to distinct
$(u,W)$ pairs are again assumed to be independent. Note that we allow
the same node $u$ to be source for several multicast groups $W$, and the
same subset $W$ of nodes to be multicast group for several sources $u$.
The \emph{multicast capacity region} $\Lambdamc(n)\subset\Rp^{n\times
2^n}$ is the closure of the collection of all achievable multicast
traffic matrices $\lambdamc\in\Rp^{n\times 2^n}$. Observe that unicast
traffic is a special case of multicast traffic, and hence $\Lambdauc(n)$
is a $\R^{n\times n}$-dimensional ``slice'' of the $\R^{n\times
2^n}$-dimensional region $\Lambdamc(n)$. 

The next example illustrates the definitions of unicast and multicast
traffic.
\begin{example}
    Consider $n=4$ and $V(n) = \{v_i\}_{i=1}^4$. Assume node $v_1$ wants
    to transmit a message $m_{1,2}$ to node $v_2$ at a rate of $1$ bit
    per second, and a message $m_{1,3}$ to node $v_3$ at rate $2$ bits
    per second. Node $v_2$ wants to transmit a message $m_{2,3}$ at rate
    $3$ bits per second to node $v_3$. The messages
    $\{m_{1,2},m_{1,3},m_{2,3}\}$ are assumed to be independent. This
    traffic requirement can be described by a unicast traffic matrix
    $\lambdauc\in\Rp^{4\times 4}$ with $\lambdauc_{v_1,v_2}\defeq 1$,
    $\lambdauc_{v_1,v_3}\defeq 2$, $\lambdauc_{v_2,v_3}\defeq 3$, and
    $\lambdauc_{u,w}\defeq 0$ for all other $(u,w)$ pairs. Note that
    node $v_1$ is source for $v_2$ and $v_3$, and that node $v_3$ is
    destination for $v_1$ and $v_2$. Note also that node $v_4$ is
    neither a source nor a destination for any communication pair, and
    can hence be understood as a helper node.

    Assume now node $v_1$ wants to transmit the same message
    $m_{1,\{3,4\}}$ to both $v_3$ and $v_4$ at rate $1$ bit per second,
    and a private message $m_{1,\{3\}}$ to only node $v_3$ at rate $2$
    bits per second. Moreover, node $v_2$ wants to transmit the same
    message $m_{2,\{3,4\}}$ to both $v_3$ and $v_4$ at rate $3$ bits per
    second. The messages $\{m_{1,\{3,4\}}, m_{1,\{3\}}, m_{2,\{3,4\}}\}$
    are assumed to be independent. This traffic requirement can be
    described by a multicast traffic matrix $\lambdamc\in\Rp^{4\times
    16}$ with $\lambdamc_{v_1,\{v_3,v_4\}}\defeq 1$,
    $\lambdamc_{v_1,\{v_3\}}\defeq 2$,
    $\lambdamc_{v_2,\{v_3,v_4\}}\defeq 3$, and $\lambdamc_{u,W}\defeq 0$
    for all other $(u,W)$ pairs. Note that $v_1$ is source for two
    multicast groups $\{v_3,v_4\}$ and $\{v_3\}$, and that $\{v_3,v_4\}$
    is multicast group for two sources $v_1$ and $v_2$. 
\end{example}

Throughout, we denote by $\log$ and $\ln$ the logarithms with respect to
base $2$ and $e$, respectively.  To simplify notation, we suppress the
dependence on $n$ within proofs whenever this dependence is clear from
the context.

\section{Main Results}
\label{sec:main}

We now present the main results of this paper. Section
\ref{sec:main_unicast} provides a scaling characterization of the
unicast capacity region $\Lambdauc(n)$, and Section
\ref{sec:main_multicast} provides a scaling characterization of the
multicast capacity region $\Lambdamc(n)$ of a dense wireless network.
Section~\ref{sec:main_examples} contains example scenarios illustrating
applications of the main theorems.

\subsection{Unicast Traffic}
\label{sec:main_unicast}

Define 
\begin{align*}
    \hLambdauc(n) \defeq
    \bigg\{\lambdauc\in\Rp^{n\times n}:
    \sum_{w\neq u}\lambdauc_{u,w} \leq 1 \ \forall u \in V(n), 
    \ \sum_{u\neq w}\lambdauc_{u,w} \leq 1 \ \forall w \in V(n)
    \bigg\}.
\end{align*}
$\hLambdauc(n)$ is the collection of all unicast traffic matrices
$\lambdauc\in\Rp^{n\times n}$ such that for every node $u$ in the network
the total traffic
\begin{equation*}
    \sum_{w\neq u}\lambdauc_{u,w}
\end{equation*}
from $u$ is less than one, and such that for every node $w$ in the
network the total traffic
\begin{equation*}
    \sum_{u\neq w}\lambdauc_{u,w}
\end{equation*}
to $w$ is less than one. 

The next theorem shows that $\hLambdauc(n)$ is a tight
approximation of the unicast capacity region $\Lambdauc(n)$ of the
wireless network.
\begin{theorem}
    \label{thm:unicast}
    For all $\alpha\geq 2$, $n\geq 9$, and node placement $V(n)$ with
    minimum node separation $r_{\min}(n)n^{-1/2}$,
    \begin{equation*} 
        2^{-\alpha/2}\hLambdauc(n)
        \subset \Lambdauc(n)
        \subset \log\big(n^{2+\alpha/2}r_{\min}^{-\alpha}(n)\big)\hLambdauc(n).
    \end{equation*}
\end{theorem}

Assuming that $r_{\min}(n)$ decays no faster than polynomial in $n$ (see
the discussion in Section~\ref{sec:model}), Theorem~\ref{thm:unicast}
states that $\hLambdauc(n)$ approximates $\Lambdauc(n)$ up to a factor
$O(\log(n))$. In other words, $\hLambdauc(n)$ provides a scaling
characterization of the unicast capacity region $\Lambdauc(n)$. This
scaling characterization is considerably more general than the standard
scaling results, in that it holds for \emph{any} node placement and
provides information on the \emph{entire} $n\times n$-dimensional
unicast capacity region (see Fig.~\ref{fig:unicast_approx}). In
particular, define
\begin{equation*}
    \rho_{\lambdauc}^{\star}(n)
    \defeq \max\{\rho: \rho\lambdauc\in\Lambdauc(n)\}
\end{equation*}
to be the largest multiple $\rho$ such that $\rho\lambdauc$ is
achievable. Then, for any \emph{arbitrary} node placement $V(n)$ and
\emph{arbitrary} unicast traffic matrix $\lambdauc\in\Rp^{n\times n}$,
Theorem~\ref{thm:unicast} determines $\rho_{\lambdauc}^{\star}(n)$ up to a
multiplicative gap of order $O(\log(n))$ uniform in $\lambdauc$. This
contrasts with the standard scaling results, which provide information
on $\rho_{\lambdauc}^{\star}(n)$ only for a \emph{uniform random} node placement
$V(n)$ and a \emph{uniform random} unicast traffic matrix $\lambdauc$
(constructed by pairing nodes randomly into $n$ source-destination pairs with uniform
rate).

\begin{figure}[tbp]
    \begin{center}
        \input{figs/unicast_approx.pstex_t}
    \end{center}

    \caption{
    The set $\hLambdauc(n)$ approximates the unicast capacity region
    $\Lambdauc(n)$ of the wireless network in the sense that
    $b_1(n)\hLambdauc(n)$, with $b_1(n)=2^{-\alpha/2}$, provides an
    inner bound to $\Lambdauc(n)$ and $b_2(n)\hLambdauc(n)$, with
    $b_2(n) = \log\big(n^{2+\alpha/2}r_{\min}^{-\alpha}(n)\big)$, provides
    an outer bound to $\Lambdauc(n)$. The figure shows two dimensions
    (namely $\lambdauc_{1,2}$ and $\lambdauc_{2,1}$) of the
    $n\times n$-dimensional set $\Lambdauc(n)$. 
    }

    \label{fig:unicast_approx}
\end{figure}

Theorem~\ref{thm:unicast} also reveals that the unicast capacity region
of a dense wireless network has a rather simple structure in that it can
be approximated up to a factor $O(\log(n))$ by an intersection of $2n$
half-spaces. Each of these half-spaces corresponds to a \emph{cut} in the
wireless network, bounding the total rate across this cut. While there
are $2^n$ such cuts in the network, Theorem~\ref{thm:unicast} implies
that only a small fraction of them are of asymptotic relevance.  From
the definition of $\hLambdauc(n)$, these are precisely the
cuts involving just a single node (with traffic flowing either into or
out of that node).

\subsection{Multicast Traffic}
\label{sec:main_multicast}

Let
\begin{align}
    \label{eq:lambdamc}
    \hLambdamc(n) \defeq
    \bigg\{\lambdamc \in\Rp^{n\times 2^n}:
    \sum_{\substack{W\subset V(n): \\ W\setminus \{u\}\neq\emptyset}}
    \lambdamc_{u,W} \leq 1 \ \forall u \in V(n), 
    \ \sum_{u\neq w}\sum_{\substack{W\subset V(n): \\ w\in W}}
    \lambdamc_{u,W} \leq 1 \ \forall w \in V(n)
    \bigg\}.
\end{align}
Similarly to $\hLambdauc(n)$ defined in Section~\ref{sec:main_unicast},
the region $\hLambdamc(n)$ is the collection of multicast traffic
matrices $\lambdamc\in\Rp^{n\times 2^n}$ such that for every node $u$ in
the network the total traffic
\begin{equation*}
    \sum_{\substack{W\subset V(n): \\ W\setminus \{u\}\neq\emptyset}}
    \lambdamc_{u,W}
\end{equation*}
from $u$ is less than one, and such that for every node $w$ in the
network the total traffic
\begin{equation*}
    \sum_{u\neq w}\sum_{\substack{W\subset V(n): \\ w\in W}}
    \lambdamc_{u,W}
\end{equation*}
to $w$ is less than one. 

The next theorem shows that $\hLambdamc(n)$ is a tight approximation of the
multicast capacity region $\Lambdamc(n)$ of the wireless network.
\begin{theorem}
    \label{thm:multicast}
    For all $\alpha\geq 2$, $n\geq 9$, and node placement $V(n)$ with
    minimum node separation $r_{\min}(n)n^{-1/2}$,
    \begin{equation*}
        2^{-1-\alpha/2}\hLambdamc(n)
        \subset \Lambdamc(n)
        \subset \log\big(n^{2+\alpha/2}r_{\min}^{-\alpha}(n)\big)\hLambdamc(n).
    \end{equation*}
\end{theorem}

Assuming as before that $r_{\min}(n)$ decays no faster than polynomial
in $n$, Theorem~\ref{thm:multicast} asserts that $\hLambdamc(n)$
approximates $\Lambdamc(n)$ up to a factor $O(\log(n))$. In other words,
as in the unicast case, we obtain a scaling characterization of the
multicast capacity region $\Lambdamc(n)$. Again, this scaling
characterization is considerably more general than standard scaling
results, in that it holds for \emph{any} node placement and provides
information about the \emph{entire} $n\times 2^n$-dimensional multicast
capacity region $\Lambdamc(n)$.  Define, as for unicast traffic
matrices,
\begin{equation*}
    \rho_{\lambdamc}^{\star}(n)
    \defeq \max\{\rho: \rho\lambdamc\in\Lambdamc(n)\}
\end{equation*}
to be the largest multiple $\rho$ such that $\rho\lambdamc$ is
achievable. Then Theorem~\ref{thm:multicast} allows, for any
\emph{arbitrary} node placement $V(n)$ and \emph{arbitrary} multicast
traffic matrix $\lambdamc\in\Rp^{n\times 2^n}$, to determine
$\rho_{\lambdamc}^{\star}(n)$ up to a multiplicative gap of order
$O(\log(n))$ uniform in $\lambdamc$. In particular, no probabilistic
assumptions about the structure of $V(n)$ or $\lambdamc$ are necessary.

As with $\Lambdauc(n)$, Theorem~\ref{thm:multicast} implies that the
multicast capacity region of a dense wireless network is approximated up
to a factor $O(\log(n))$ by an intersection of $2n$ half spaces.  In
other words, we are approximating a region of dimension $n\times 2^n$
(i.e., exponentially big in $n$) through only a linear number of
inequalities. As in the case of unicast traffic, each of these
inequalities corresponds to a cut in the wireless network, and it is
again the cuts involving just a single node that are asymptotically
relevant.

\subsection{Examples}
\label{sec:main_examples}

This section contains several examples illustrating various aspects of
the capacity regions $\Lambdauc(n), \Lambdamc(n)$ and their
approximations $\hLambdauc(n), \hLambdamc(n)$. Example \ref{eg:sd}
compares the scaling laws obtained in this paper with the ones obtained
using hierarchical cooperation as proposed in \cite{ozgur07b}. Example
\ref{eg:symmetry} discusses symmetry properties of $\Lambdauc(n)$ and
$\Lambdamc(n)$. Example \ref{eg:outer} provides a traffic pattern
showing that the outer bounds in Theorems \ref{thm:unicast} and
\ref{thm:multicast} are tight up to a constant factor.

\newpage

\begin{example}
    \label{eg:sd}
    (\emph{Random source-destination pairing})

    Consider a random node placement $V(n)$ with every node placed
    independently and uniformly at random on $A$.  Assume we pair each
    node $u\in V(n)$ with a node $w\in V(n)\setminus\{u\}$ chosen
    independently and uniformly at random. Denote by $\{u_i,w_i\}$ the
    resulting $n$ source-destination pairs. Note that each node is
    source exactly once and destination on average once. Each source
    $u_i$ wants to transmit an independent message to $w_i$ at rate
    $\rho(n)$ (depending on $n$, but not on $i$). The question is to
    determine $\rho^{\star}(n)$, the largest achievable value of
    $\rho(n)$. This question was considered in \cite{ozgur07b}, where it
    was shown that, with probability $1-o(1)$ as $n\to\infty$ and for
    every $\varepsilon> 0$,
    \begin{equation}
        \label{eq:example1a}
        \Omega(n^{-\varepsilon})
        \leq \rho^{\star}(n)
        \leq O(n^{\varepsilon}).
    \end{equation}
    The lower bound is achieved by a hierarchical cooperation scheme,
    and we denote its rate by $\rho_{\mathrm{HC}}(n)$.

    We now show that using the results presented in this paper these
    bounds on $\rho^{\star}(n)$ can be significantly sharpened. Set
    $\lambdauc_{u_i,w_i} \defeq 1$ for $i\in\{1,\ldots,n\}$ and
    $\lambdauc_{u,w}\defeq 0,$ for all other entries of $\lambdauc$.
    $\rho^{\star}(n)$ is then given by
    \begin{equation*}
        \rho^{\star}(n) 
        = \max\{\rho:\rho\lambdauc\in\Lambdauc(n)\}.
    \end{equation*}
    Setting
    \begin{equation*}
        \hat{\rho}^{\star}(n) 
        \defeq \max\{\hat{\rho}:\hat{\rho}\lambdauc\in\hLambdauc(n)\},
    \end{equation*}
    we obtain from Theorem~\ref{thm:unicast} that
    \begin{equation}
        \label{eq:example1b}
        2^{-\alpha/2}\hat{\rho}^{\star}(n)
        \leq \rho^{\star}(n)
        \leq \log\big(n^{2+\alpha/2}r_{\min}^{-\alpha}(n)\big)\hat{\rho}^{\star}(n).
    \end{equation}

    It remains to evaluate $\hat{\rho}^{\star}(n)$. By construction of
    $\lambdauc$, we have
    \begin{equation*}
        \max_{u\in V(n)}\sum_{w\neq u}\lambdauc_{u,w} = 1.
    \end{equation*}
    Moreover, by~\cite{raab98},
    \begin{equation*}
        \Pp\bigg(
        \frac{1}{2} \leq
        \frac{\ln\ln(n)}{\ln(n)}\max_{w\in V(n)} \sum_{u\neq w} \lambdauc_{u,w} 
        \leq 2
        \bigg)
        \geq 1-o(1).
    \end{equation*}
    Using the definition of $\hLambdauc(n)$, this yields that
    \begin{equation}
        \label{eq:rhohat3}
        \frac{\ln\ln(n)}{2\ln(n)}\leq 
        \hat{\rho}^{\star}(n) 
        \leq \frac{2\ln\ln(n)}{\ln(n)}
    \end{equation}
    with high probability. 

    Recall that the minimum distance between nodes is
    $r_{\min}(n)n^{-1/2}$, and that, for a random node placement,
    $r_{\min}(n) \geq n^{-1}$ with high probability as $n\to\infty$
    (see, e.g., \cite[Theorem 3.1]{ozgur07b}). Hence
    \eqref{eq:example1b} and \eqref{eq:rhohat3} show that that for
    random node placement and random source-destination pairing
    \begin{equation}
        \label{eq:example1c}
        2^{-1-\alpha/2}\frac{\ln\ln(n)}{\ln(n)}
        \leq \rho^{\star}(n)
        \leq (4+3\alpha)\log(e)\ln\ln(n)
    \end{equation}
    with probability $1-o(1)$ as $n\to\infty$. The lower bound is
    achieved using a communication scheme presented in Section
    \ref{sec:schemes_unicast} based on interference alignment, and we
    denote its rate by $\rho_{\mathrm{IA}}(n)$.
    
    Comparing \eqref{eq:example1c} and \eqref{eq:example1a}, we see that
    the scaling law obtained here is significantly sharper, namely up to
    a factor $O(\log(n))$ here as opposed to a factor $O(n^{\varepsilon})$ for
    any $\varepsilon>0$ in \cite{ozgur07b}. Moreover,
    \eqref{eq:example1c} provides good estimates for any value of $n$,
    whereas \eqref{eq:example1a} is only valid for large values of $n$,
    with a pre-constant in $O(n^{\varepsilon})$ that increases rapidly
    as $\varepsilon\to 0$ (see \cite{ghaderi09,xie09} for a detailed
    discussion on the dependence of the pre-constant on $\varepsilon$).
    For a numerical example, Table \ref{tbl:comp} compares per-node
    rates $\rho_{\mathrm{HC}}(n)$ of the hierarchical cooperation scheme
    of \cite{ozgur07b} (more precisely, an upper bound to it, with
    optimized parameters as analyzed in \cite{ghaderi09}) with the
    per-node rates $\rho_{\mathrm{IA}}(n)$ obtained through interference
    alignment as proposed in this paper. For the numerical example, we
    choose $\alpha=4$.

    \begin{table}[tbp]
        \renewcommand\arraystretch{1.2}
        \caption{Comparison of $\rho_{\text{HC}}(n)$ and
        $\rho_{\text{IA}}(n)$ (in bits per channel use) from Example \ref{eg:sd}.}

        \begin{center}
            \begin{tabular}{r| r@{.}l r@{.}l r@{.}l r@{.}l}
                & 
                \multicolumn{2}{c}{$n=10^2$} &
                \multicolumn{2}{c}{$n=10^3$} &
                \multicolumn{2}{c}{$n=10^4$} &
                \multicolumn{2}{c}{$n=10^5$} \\
                \hline
                $\rho_{\mathrm{HC}}(n)$ & 
                0&0017 &
                0&00047 & 
                0&00017 & 
                0&000070 \\
                $\rho_{\mathrm{IA}}(n)$ & 
                0&042 & 
                0&035 & 
                0&030 &
                0&027
            \end{tabular}
        \end{center}
        \renewcommand\arraystretch{1.0}

        \label{tbl:comp} 
    \end{table}

    We point out that the per-node rate $\rho_{\mathrm{IA}}(n)$ decreases
    as the number of nodes $n$ increases only because of the random
    source-destination pairing. In fact, if the nodes $\{u_i,w_i\}$ are
    paired such that each node is source and destination exactly once,
    then the interference alignment based scheme achieves a per-node
    rate $\rho_{\mathrm{IA}}\geq 2^{-\alpha/2}$, i.e., the
    per-node rate does not decay to zero as $n\to\infty$. 
\end{example}

\begin{example}
    \label{eg:symmetry}
    (\emph{Symmetry of $\Lambdauc(n)$ and $\Lambdamc(n)$})

    Theorems \ref{thm:unicast} and \ref{thm:multicast} provide some
    insight into (approximate) symmetry properties of the unicast and
    multicast capacity regions $\Lambdauc(n)$ and $\Lambdamc(n)$.
    Indeed, their approximations $\hLambdauc(n)$ and $\hLambdamc(n)$ are
    invariant with respect to node positions (and hence, in particular,
    also invariant under permutation of nodes). 
    
    More precisely, consider a unicast traffic matrix
    $\lambdauc\in\Rp^{n\times n}$. For a permutation $\pi$ of the nodes
    $V(n)$ set
    \begin{equation*}
        \tlambdauc_{u,w} \defeq \lambdauc_{\pi(u),\pi(w)}.
    \end{equation*}
    Then $\lambdauc\in\hLambdauc(n)$ if and only if
    $\tlambdauc\in\hLambdauc(n)$. Hence Theorem~\ref{thm:unicast} yields
    that if $\lambdauc\in\Lambdauc(n)$, then
    \begin{equation*}
        2^{-\alpha/2}\log^{-1}\big(n^{2+\alpha/2}r_{\min}^{-\alpha}(n)\big)\tlambdauc
        \in\Lambdauc(n).
    \end{equation*}
    Similarly, let $\lambdamc\in\Rp^{n\times 2^n}$ be
    a multicast traffic matrix, and define 
    \begin{equation*}
        \tlambdamc_{u,W} \defeq \lambdamc_{\pi(u),\pi(W)},
    \end{equation*}
    where, for $W\subset V(n)$, $\pi(W)\defeq \{\pi(w): w\in W\}$. 
    Theorem~\ref{thm:multicast} implies that if
    $\lambdamc\in\Lambdamc(n)$, then
    \begin{equation*}
        2^{-1-\alpha/2}\log^{-1}\big(n^{2+\alpha/2}r_{\min}^{-\alpha}(n)\big)\tlambdamc
        \in\Lambdamc(n).
    \end{equation*}

    In other words, the location of the nodes in a dense wireless
    network (with $r_{\min}(n)$ decaying at most polynomially in $n$)
    affects achievable rates at most up to a factor $O(\log(n))$. 
    This contrasts with the behavior of extended wireless networks,
    where node locations crucially affect achievable rates 
    \cite{niesen09}.
\end{example}

\begin{example}
    \label{eg:outer}
    (\emph{Tightness of outer bounds})

    We now argue that the outer bounds in Theorems
    \ref{thm:unicast} and \ref{thm:multicast} are tight up to a constant
    factor in the following sense.  There exists a constant $K>0$
    such that for every $n$ we can find traffic matrices
    $\lambdauc$ and $\lambdamc$ on the boundary of the outer bound in
    Theorems \ref{thm:unicast} and \ref{thm:multicast} such that
    $K\lambdauc\in\Lambdauc(n)$ and $K\lambdamc\in\Lambdamc(n)$. Or,
    more succinctly, there exists a constant $K>0$ such that
    \begin{align*}
        \Lambdauc(n)\setminus 
        K\log\big(n^{2+\alpha/2}r_{\min}^{-\alpha}(n)\big)\hLambdauc(n) & \neq \emptyset, \\
        \Lambdamc(n)\setminus 
        K\log\big(n^{2+\alpha/2}r_{\min}^{-\alpha}(n)\big)\hLambdamc(n) & \neq \emptyset.
    \end{align*}
    This shows that the $O(\log(n))$ gap between the
    inner and outer bounds in Theorems \ref{thm:unicast} and
    \ref{thm:multicast} is due to the use of the interference alignment
    scheme to prove the inner bound, and that to further decrease this
    gap a different achievable scheme has to be considered. Throughout
    this example, we assume $r_{\min}(n) > n^{-\kappa}$ for some
    constant $\kappa\geq 0$. 

    Choose a node $w^{\star}\in V(n)$, and let, for each $u,w\in V(n)$,
    \begin{equation*}
        \lambdauc_{u,w} \defeq
        \begin{cases}
            \frac{1}{n-1} & \text{if $w=w^{\star}$}, \\
            0 & \text{otherwise}.
        \end{cases}
    \end{equation*}
    Note that $\lambdauc\in\hLambdauc(n)$. Under this traffic matrix
    $\lambdauc$, each node $u\in V(n)$ has an independent message for a
    common destination node $w^{\star}$. 
    
    If we ignore the received signals at
    all nodes $v\neq w^{\star}$ and transmit no signal at $w^{\star}$, we transform
    the wireless network into a multiple access channel with $n-1$
    users. Since $r_{u,w^{\star}}\leq\sqrt{2}$ for any $u\in V(n)$, each node
    $u\in V(n)\setminus\{w^{\star}\}$ can reduce its power such that the
    received power at node $w^{\star}$ is equal to $2^{-\alpha/2}$. In this
    symmetric setting, the equal rate point of the capacity region of the
    multiple access channel has maximal sum rate, and hence each node $u\in
    V(n)\setminus\{w^{\star}\}$ can reliably transmit its message to $w^{\star}$ at
    a per-node rate of 
    \begin{align*}
        \frac{1}{n-1}\log(1+ (n-1) 2^{-\alpha/2}) 
        & \geq \frac{1}{n-1}\log(n2^{-\alpha/2}) \\
        & = \frac{1}{n-1}\Big(1-\frac{\alpha}{2\log(n)}\Big)\log(n).
    \end{align*}
    Thus, for $n> 2^{\alpha}$,
    \begin{equation}
        \label{eq:eg_outer1}
        \frac{1}{2}\log(n)\lambdauc \in\Lambdauc(n).
    \end{equation}

    On the other hand, using the assumption $r_{\min}(n) >
    n^{-\kappa}$,
    \begin{equation*}
        \log\big(n^{2+\alpha/2}r_{\min}^{-\alpha}(n)\big)
        < \big(2+\alpha(1/2+\kappa)\big)\log(n),
    \end{equation*}
    and hence
    \begin{equation}
        \label{eq:eg_outer2}
        \big(2+\alpha(1/2+\kappa)\big)\log(n)\lambdauc
        \notin\log\big(n^{2+\alpha/2}r_{\min}^{-\alpha}(n)\big)\hLambdauc(n).
    \end{equation}

    Therefore, setting
    \begin{equation*}
        K \defeq \big(4+\alpha(1+2\kappa)\big)^{-1} > 0,
    \end{equation*}
    we obtain from \eqref{eq:eg_outer1} and \eqref{eq:eg_outer2} that
    \begin{equation*}
        \Lambdauc(n)\setminus K\log\big(n^{2+\alpha/2}r_{\min}^{-\alpha}(n)\big)\hLambdauc(n)
        \neq \emptyset.
    \end{equation*}
    In words, at least along one direction in $\R^{n\times n}$, the
    outer bound in Theorem~\ref{thm:unicast} is loose by at most a
    constant factor. 
    
    Since $\Lambdauc(n)$ is a $n\times n$-dimensional ``slice'' of the
    $n\times 2^n$-dimensional region $\Lambdamc(n)$, the same result
    follows for $\Lambdamc(n)$ as well.
\end{example}

\section{Communication Schemes}
\label{sec:schemes}

This section describes the communication schemes achieving the inner bounds
in Theorems \ref{thm:unicast} and \ref{thm:multicast}. Both schemes use the
idea of interference alignment as a building block, which is recalled in
Section~\ref{sec:schemes_alignment}. The communication scheme for unicast
traffic is introduced in Section~\ref{sec:schemes_unicast} and the scheme for
multicast traffic in Section~\ref{sec:schemes_multicast}.

\subsection{Interference Alignment}
\label{sec:schemes_alignment}

Interference alignment is a technique introduced recently in
\cite{cadambe08, maddah-ali08}. The technique is best illustrated with
an example taken from \cite{nazer09}. Assume we pair the nodes $V(n)$
into source-destination pairs $\{u_i,w_i\}_{i=1}^{n}$ such that each
node in $V(n)$ is source and destination exactly once. Consider the
channel gains $\{h_{u_i,w_j}[t_1]\}_{i,j}$ and
$\{h_{u_i,w_j}[t_2]\}_{i,j}$ for two different times $t_1$ and
$t_2$. Assume we could choose $t_1$ and $t_2$ such that
$h_{u_i,w_i}[t_1] = h_{u_i,w_i}[t_2]$ and $h_{u_i,w_j}[t_1] =
-h_{u_i,w_j}[t_2]$ for all $i\neq j$. By adding up the received symbols
$y_{w_i}[t_1]$ and $y_{w_i}[t_2]$, destination node $w_i$ obtains
\begin{equation*}
    y_{w_i}[t_1]+y_{w_i}[t_2] 
    = h_{u_i,w_i}[t_1](x_{u_i}[t_1]+x_{u_i}[t_2])+z_{w_i}[t_1]+z_{w_i}[t_2].
\end{equation*}
Thus, by sending the same symbol twice (i.e.,
$x_{u_i}[t_1]=x_{u_i}[t_2]$), every source node $u_i$ is able to
communicate with its destination node $w_i$ at essentially half the rate
possible without any interference from other nodes.

Using this idea and the symmetry and ergodicity of the distribution of
the channel gains, the following result is shown in \cite{nazer09}.
\begin{theorem}
    \label{thm:alignment}
    For any source-destination pairing $\{u_i,w_i,\}_{i=1}^{n}$ such
    that $u_i\neq u_j$ and $w_i\neq w_j$ for $i\neq j$, the rates
    \begin{equation*}
        \lambdauc_{u_i,w_j} =
        \begin{cases}
            \frac{1}{2}\log(1+2\abs{h_{u_i,w_i}}^2) & \text{if $i=j$}, \\
            0 & \text{otherwise},
        \end{cases}
    \end{equation*}
    are achievable, i.e., $\lambdauc\in\Lambdauc(n)$.
\end{theorem}
For a source-destination pairing $\{u_i,w_i,\}_{i=1}^{n}$ as in Theorem
\ref{thm:alignment}, construct a matrix $S\in\Rp^{n\times n}$ such that
\begin{equation*}
    S_{u_i,w_j} =
    \begin{cases}
        1 & \text{if $i=j$}, \\
        0 & \text{otherwise}.
    \end{cases}
\end{equation*}
Note that $S$ is a permutation matrix, and we will call such a traffic
pattern a \emph{permutation traffic}. Using $r_{u_i,w_i}\leq \sqrt{2}$
and $\alpha\geq 2$,
\begin{equation*}
    \frac{1}{2}\log(1+2r_{u_i,w_i}^{-\alpha})
    \geq \frac{1}{2}\log(1+2^{1-\alpha/2})
    \geq 2^{-\alpha/2},
\end{equation*}
and hence Theorem~\ref{thm:alignment} provides an achievable scheme
showing that $2^{-\alpha/2}S\in\Lambdauc(n)$. In other words,
Theorem~\ref{thm:alignment} shows that, for every permutation traffic, a
per-node rate of $2^{-\alpha/2}$ is achievable. In the next two
sections, we will use this communication scheme for permutation traffic
as a building block to construct communication schemes for general
unicast and multicast traffic.

\subsection{Communication Scheme for Unicast Traffic}
\label{sec:schemes_unicast}

Consider a general unicast traffic matrix $\lambdauc\in\Rp^{n\times n}$.
If $\lambdauc$ happens to be a scalar multiple of a permutation matrix,
then Theorem~\ref{thm:alignment} provides us with an achievable scheme
to transmit according to $\lambdauc$. In order to apply Theorem
\ref{thm:alignment} for general $\lambdauc$, we  need to
\emph{schedule} transmissions into several slots such that in each slot
transmission occurs according to a permutation traffic. This
transforms the original problem of communicating over a
wireless network into a problem of scheduling over a switch with $n$
input and $n$ output ports and traffic requirement $\lambdauc$. 

This problem has been widely studied in the literature.  In particular,
using a result from von Neumann \cite{vonneumann53} and Birkhoff
\cite{birkhoff46} (see also \cite{chang00} for the application to
switches) it can be shown that for any $\lambdauc\in\hLambdauc(n)$ there
exist a collection of \emph{schedules} $\{S_i\}$ (essentially
permutation matrices, see the proof in Section~\ref{sec:proofs_unicast}
for the details) and nonnegative weights $\{\omega_i\}$ summing to one
such that
\begin{equation*}
    \sum_{i}\omega_i S_i = \lambdauc.
\end{equation*}

This suggests the following communication scheme.  Split time into slots
according to the weights $\{\omega_i\}$. In the slot corresponding to
$\omega_i$, send traffic over the wireless network using interference
alignment for the schedule $S_i$. In other words, we time share between
the different schedules $\{S_i\}$ according to the weights
$\{\omega_i\}$.

We analyze this communication scheme in more detail in Section
\ref{sec:proofs_unicast}. In particular, we show that it achieves any
point in $2^{-\alpha/2}\hLambdauc(n)$. Combined with a matching outer
bound, we show that this scheme is optimal for any unicast traffic
pattern up to a factor
$2^{\alpha/2}\log\big(n^{2+\alpha/2}r_{\min}^{-\alpha}(n)\big)$.

Recall from Example~\ref{eg:symmetry} that the capacity region is
approximately symmetric with respect to permutation of the traffic
matrix. This implies that the rate achievable for any permutation
traffic is approximately the same. While the decomposition of the
traffic matrix $\lambdauc$ into schedules $\{S_i\}$ is not unique, this
invariance suggests that it does not matter too much which decomposition
is chosen. The situation is different for Rayleigh fading (as opposed to
phase fading considered here), where different decompositions can be
used for opportunistic communication. This approach is explored in
detail in Section~\ref{sec:discussion_rayleigh}.

\subsection{Communication Scheme for Multicast Traffic}
\label{sec:schemes_multicast}

We now turn to multicast traffic. Given the achievable scheme presented
for unicast traffic in Section~\ref{sec:schemes_unicast} reducing the
problem of communication over a wireless network to that of scheduling
over a switch, it is tempting to try the same approach for multicast
traffic as well. Unfortunately, scheduling of multicast traffic over
switches is considerably more difficult than the corresponding unicast
version (see, for example, \cite{marsan03} for converse results showing
the infeasibility of multicast scheduling over switches with finite
speedup). We therefore adopt a different approach here. The proposed
communication scheme is reminiscent of the two-phase routing scheme of
Valiant and Brebner \cite{valiant81}.

Consider a source node $u\in V(n)$ that wants to multicast a message to
destination group $W\subset V(n)$. The proposed communication scheme
operates in two phases. In the first phase, the node $u$ splits its
message into $n$ parts of equal length. It then sends one (distinct)
part over the wireless network to each node in $V(n)$. Thus, after the
first phase, each node in $V(n)$ has access to a distinct fraction $1/n$
of the original message. In the second phase, each node in $V(n)$ sends
its message parts to all the nodes in $W$. Thus, at the end of the
second phase, each node in $W$ can reconstruct the entire message. All
pairs $(u,W)$ operate simultaneously within each phase, and contention
within the phases is resolved by appropriate scheduling (see the proof
in Section~\ref{sec:proofs_multicast} for the details).

\begin{figure}[tbp]
    \begin{center}
        \scalebox{0.889}{
        \input{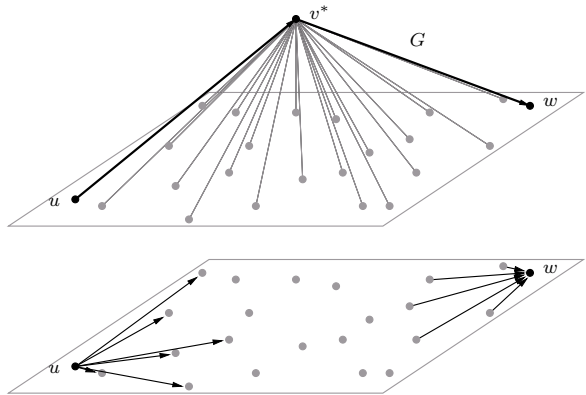}
        }
    \end{center}

    \caption{Construction of the ``star'' graph $G$, and parts of the
    corresponding induced transmissions in the underlying wireless
    network for communication between $u$ and $w$.}

    \label{fig:star}
\end{figure}

A different way to look at this proposed communication scheme is
as follows.  Consider the $n$ nodes in $V(n)$, and construct a graph
$G=(V_G, E_G)$ with $V_G \defeq V(n)\cup\{v^{\star}\}$ for some additional
node $v^{\star}\notin V(n)$ and with $(u,v) \in E_G$ if either $u=v^{\star}$ or
$v=v^{\star}$. In other words, $G$ is a ``star'' graph with central node $v^{\star}$
(see Fig.~\ref{fig:star}). We assign to each edge $e\in E_G$ an edge
capacity of one. The proposed communication scheme for the wireless
network can then be understood as a two layer architecture, consisting
of a \emph{physical layer} and a \emph{network layer}. The physical
layer implements the graph abstraction $G$, and the network layer routes
data over $G$.

In Section~\ref{sec:proofs_multicast}, we show that the set of rates
$\Lambdamc_G(n)$ that can be routed over $G$ contains $\hLambdamc(n)$.
We then argue that if $\lambdamc\in 2^{-1-\alpha/2}\Lambdamc_G(n)$, then
$\lambdamc\in\Lambdamc(n)$, i.e., if messages can be routed over the
graph $G$ at rates $\lambdamc$, then almost the same rates are
achievable in the wireless network.  Combining this with a matching
outer bound, we show that the proposed communication scheme is optimal
for any multicast traffic pattern up to a factor
$2^{1+\alpha/2}\log\big(n^{2+\alpha/2}r_{\min}^{-\alpha}(n)\big)$.

\newpage

\section{Proofs}
\label{sec:proofs}

This section contains the proofs of Theorem~\ref{thm:unicast} (in Section
\ref{sec:proofs_unicast}) and Theorem~\ref{thm:multicast} (in Section
\ref{sec:proofs_multicast}).

\subsection{Proof of Theorem~\ref{thm:unicast}}
\label{sec:proofs_unicast}

We start with the proof of the outer bound in Theorem~\ref{thm:unicast}.
For subsets $S_1,S_2\subset V$, $S_1\cap S_2 = \emptyset$, denote by
$C(S_1,S_2)$ the capacity of the multiple-input multiple-output (MIMO)
channel between nodes in $S_1$ and nodes in $S_2$.
Applying the cut-set bound \cite[Theorem 14.10.1]{cover91} to the
sets $S_1\defeq \{w\}^c$, $S_2 \defeq \{w\}$, we obtain
\begin{equation*}
    \sum_{u\neq w}\lambdauc_{u,w} \leq C(\{w\}^c,\{w\}).
\end{equation*}
$C(\{w\}^c,\{w\})$ is upper bounded by relaxing the individual
power constraints at each node to a sum-power constraint of $n-1$.
This yields
\begin{align*}
    C(\{w\}^c,\{w\}) 
    & \leq \log\Big(1+ (n-1){ \textstyle \sum_{u\neq w} } \abs{h_{u,w}}^2\Big) \\
    & = \log\Big(1+ (n-1){ \textstyle \sum_{u\neq w} } r_{u,w}^{-\alpha}\Big).
\end{align*}
Since $r_{u,w}\geq n^{-1/2}r_{\min}$, we can continue this as
\begin{align*}
    \log\Big(1+ (n-1) { \textstyle \sum_{u\neq w} } r_{u,w}^{-\alpha}\Big)
    & \leq \log\big(1+(n-1)^2 n^{\alpha/2}r_{\min}^{-\alpha}\big) \\
    & \leq \log\big(1-n^{1+\alpha/2}3^{-\alpha}+n^{2+\alpha/2}r_{\min}^{-\alpha}\big) \\
    & \leq \log\big(n^{2+\alpha/2}r_{\min}^{-\alpha}\big),
\end{align*}
where we have used that $r_{\min}\leq 3$ by \eqref{eq:rmin} and that
$n\geq 9$ by assumption.  Hence
\begin{equation}
    \label{eq:cutset2}
    \sum_{u\neq w}\lambdauc_{u,w} 
    \leq \log\big(n^{2+\alpha/2}r_{\min}^{-\alpha}\big)
\end{equation}
for all $w\in V$.  Similarly,
\begin{equation}
    \label{eq:cutset3}
    \sum_{w\neq u}\lambdauc_{u,w} 
    \leq \log\big(n^{2+\alpha/2}r_{\min}^{-\alpha}\big)
\end{equation}
for all $u\in V$.

Let $\lambdauc\in\Lambdauc$.  From
\eqref{eq:cutset2} and \eqref{eq:cutset3}, we have that 
\begin{equation*}
    \lambdauc\in
    \log\big(n^{2+\alpha/2}r_{\min}^{-\alpha}\big)\hLambdauc.
\end{equation*}
This implies
\begin{equation*}
    \Lambdauc\subset
    \log\big(n^{2+\alpha/2}r_{\min}^{-\alpha}\big)\hLambdauc,
\end{equation*}
concluding the proof of the outer bound.

We continue with the proof of the inner bound. Consider a unicast
traffic matrix $\lambdauc\in\hLambdauc$. By definition of $\hLambdauc$,
this implies that
\begin{align*}
    \sum_{w\neq u} \lambdauc_{u,w} & \leq 1 \ \forall u\in V, \\
    \sum_{u\neq w} \lambdauc_{u,w} & \leq 1 \ \forall w\in V.
\end{align*}
Moreover, we can assume without loss of generality that
$\lambdauc_{u,u}=0$ for all $u\in V$. Hence
\begin{equation}
    \label{eq:substochastic}
    \begin{aligned}
        \sum_{w} \lambdauc_{u,w} & \leq 1 \ \forall u\in V, \\
        \sum_{u} \lambdauc_{u,w} & \leq 1 \ \forall w\in V.
    \end{aligned}
\end{equation}
A matrix $\lambdauc$ satisfying the two conditions in
\eqref{eq:substochastic} is called a \emph{doubly substochastic} matrix.
If $\lambdauc$ satisfies the conditions in \eqref{eq:substochastic} with
equality, it is called a \emph{doubly stochastic} matrix. Now, by
\cite[Lemma 1]{vonneumann53}, for every doubly substochastic matrix
$\lambdauc\in\Rp^{n\times n}$, there exists a doubly stochastic matrix
$\tlambdauc\in\Rp^{n\times n}$ such that
$\lambdauc_{u,w}\leq\tlambdauc_{u,w}$ for all $u,w\in V$. If we can show
that $\tlambdauc$ is achievable, then $\lambdauc$ is achievable as well.
It suffices therefore to consider doubly stochastic traffic matrices,
and we will assume in the following that $\lambdauc$ itself is doubly
stochastic.

The set of doubly stochastic matrices of dimension $n\times n$ is convex
and compact, and hence every matrix in this set can be written as a
convex combination of its extreme points, see \cite[Corollary
18.5.1]{rockafellar96}. Now, by Birkhoff's theorem \cite[Theorem
1]{birkhoff46} (see, e.g.,  \cite[Theorem 8.7.1]{horn85} for a more
recent reference), the extreme points of the set of doubly stochastic
matrices are the permutation matrices. Hence there exists a collection
of nonnegative weights $\{\omega_i\}$ summing to one and a collection of
permutation matrices $\{S^i\}$ such that
\begin{equation}
    \label{eq:convex}
    \sum_i \omega_i S^i = \lambdauc.
\end{equation}

We time share between the different $\{S^i\}$ with weights given by
$\omega_i$. Consider now transmission of messages according to one such
permutation matrix $S^i$. Using the ergodic interference alignment
strategy proposed in \cite{nazer09} (as summarized by Theorem~\ref{thm:alignment} in
Section~\ref{sec:schemes_alignment}), each source-destination pair
$(u,w)$ such that $S^i_{u,w}=1$ can simultaneously communicate at a
per-node rate of 
\begin{equation*}
    \frac{1}{2}\log\big(1+2r_{u,w}^{-\alpha}\big)
    \geq \frac{1}{2}\log\big(1+2^{1-\alpha/2}\big) 
    \geq 2^{-\alpha/2},
\end{equation*}
where we have used that $\alpha\geq 2$ in the second inequality. Thus,
during the fraction of time corresponding to $\omega_i$, the nodes communicate
at rates $2^{-\alpha/2}S_i$.

With the time sharing described above and using \eqref{eq:convex}, this
shows that we can achieve
\begin{equation*}
    \sum_i \omega_i 2^{-\alpha/2}S_i = 2^{-\alpha/2}\lambdauc.
\end{equation*}
Therefore $2^{-\alpha/2}\lambdauc\in\Lambdauc$, which implies
\begin{equation*}
    2^{-\alpha/2}\hLambdauc\subset\Lambdauc,
\end{equation*}
proving the inner bound. \qed

\subsection{Proof of Theorem~\ref{thm:multicast}}
\label{sec:proofs_multicast}

We first prove the outer bound. Assume $\lambdamc\in\Lambdamc$. Fix a
node $u\in V$, and choose for every subset $W\subset V$ such that
$W\setminus \{u\}\neq\emptyset$ a node $\tilde{w}(W)\in
W\setminus\{u\}$. Construct the unicast traffic matrix 
\begin{equation*}
    \lambdauc_{u,w} \defeq \sum_{\substack{W\subset V: \\
    \tilde{w}(W)=w}}\lambdamc_{u,W},
\end{equation*}
for all $w\in V$, and $\lambdauc_{\tilde{u},w}\defeq 0$ for $\tilde{u}\neq u$.
Note that $\lambdamc\in\Lambdamc$ implies that $\lambdauc\in\Lambdauc$.
Indeed, we can transmit unicast traffic according to $\lambdauc$ by
using the scheme for the multicast traffic matrix $\lambdamc$ and simply
discarding the delivered messages for subset $W$ at all nodes
$W\setminus\{\tilde{w}(W)\}$. Applying Theorem~\ref{thm:unicast},
\begin{align}
    \label{eq:multicast1}
    \sum_{\substack{W\subset V: \\ W\setminus\{u\}\neq\emptyset}}\lambdamc_{u,W}
    & = \sum_{w\neq u}\sum_{\substack{W\subset V: \\ \tilde{w}(W)=w}}\lambdamc_{u,W} \nonumber\\
    & = \sum_{w\neq u}\lambdauc_{u,w} \nonumber\\
    & \leq \log\big(n^{2+\alpha/2}r_{\min}^{-\alpha}\big). 
\end{align}
Since the choice of $u$ was arbitrary, \eqref{eq:multicast1} holds for
all $u\in V$.

Fix now a node $w\in V$, and construct the unicast traffic matrix
\begin{equation*}
    \lambdauc_{u,w} \defeq \sum_{\substack{W\subset V: \\ w\in W}}\lambdamc_{u,W}
\end{equation*}
for all $u\in V$, and $\lambdauc_{u,\tilde{w}}\defeq 0$ if
$\tilde{w}\neq w$. As before, $\lambdamc\in\Lambdamc$ implies
$\lambdauc\in\Lambdauc$. Hence, by Theorem~\ref{thm:unicast}, 
\begin{align}
    \label{eq:multicast2}
    \sum_{u\neq w}\sum_{\substack{W\subset V: \\ w\in W}}\lambdamc_{u,W}
    & = \sum_{u\neq w}\lambdauc_{u,w} \nonumber\\
    & \leq \log\big(n^{2+\alpha/2}r_{\min}^{-\alpha}\big).
\end{align}
As before, the choice of $w$ was arbitrary, and hence
\eqref{eq:multicast2} holds for all $w\in V$.

Combining \eqref{eq:multicast1} and \eqref{eq:multicast2} shows that
$\lambdamc\in\Lambdamc$ implies
\begin{equation*}
    \lambdamc\in\log\big(n^{2+\alpha/2}r_{\min}^{-\alpha}\big)\hLambdamc.
\end{equation*}
Therefore
\begin{equation*}
    \Lambdamc\subset\log\big(n^{2+\alpha/2}r_{\min}^{-\alpha}\big)\hLambdamc,
\end{equation*}
proving the outer bound.

We now prove the inner bound. We construct a graph $G=(V_G,E_G)$ such
that $V\subset V_G$, i.e., the nodes in the wireless network are a
subset of the nodes in the graph $G$. We show that if messages can be
routed over $G$ at rates $\lambdamc$, then $2^{-1-\alpha/2}\lambdamc$ is
achievable over the wireless network. We then argue that $\hLambdamc$ is
a subset of the rates that are achievable by routing over $G$. Together
this will yield the desired inner bound.

The graph $G$ is a directed capacitated ``star'' graph constructed as
follows. Consider $V$ and pick an additional node $v^{\star}\notin V$. Set
\begin{align*}
    V_G & \defeq V \cup \{v^{\star}\}, \\ 
    E_G & \defeq \{(u,v): u=v^{\star} \text{ or } v=v^{\star}\}
\end{align*}
(see Fig.~\ref{fig:star} in Section~\ref{sec:schemes_multicast}).
Assign an edge capacity $c_e\defeq 1$ for all $e\in E_G$. Note that,
since $V\subset V_G$, every multicast traffic matrix
$\lambdamc\in\Rp^{n\times 2^n}$ for the wireless network is also a
multicast traffic matrix for $G$ (involving only the subset $V\subset
V_G$ of nodes as sources and destinations). Define $\Lambdamc_G$ as the
collection of such multicast traffic matrices $\lambdamc\in\Rp^{n\times
2^n}$ that are achievable via routing over $G$. 

We now argue that $\hLambdamc\subset\Lambdamc_G$. Assume
$\lambdamc\in\hLambdamc$. Since 
\begin{equation*}
    \sum_{\substack{W\subset V: \\ W\setminus \{u\}\neq\emptyset}}
    \lambdamc_{u,W} \leq 1 
\end{equation*}
for every $u\in V$, we can route all traffic $\lambdamc_{u,W}$ that is
requested at some node other than $u$ (i.e., such that
$W\setminus\{u\}\neq\emptyset$) from $u$ to the central node $v^{\star}$.
Since
\begin{equation*}
    \sum_{u\neq w}\sum_{\substack{W\subset V: \\ w\in W}}
    \lambdamc_{u,W} \leq 1
\end{equation*}
for all $w \in V$, we can route all traffic $\lambdamc_{u,W}$ that is
requested at some node $w\subset W$ from the central node $v^{\star}$ to $w$.
Together, this shows that $\lambdamc\in\Lambdamc_G$, and hence that
\begin{equation}
    \label{eq:multicast3}
    \hLambdamc\subset\Lambdamc_G.
\end{equation}

We next argue that $\Lambdamc_G\subset 2^{1+\alpha/2}\Lambdamc$. To this
end, we show that any operation on $G$ can be implemented in the
wireless network at least at a factor $2^{-1-\alpha/2}$ of the rate. For
the implementation of $G$ in the wireless network, we time share between
edges towards the central node $v^{\star}$ and from the central node. This
leads to a factor $2$ loss in rate. We implement all edges
$\{(u,v^{\star})\}_{u\in V}$ simultaneously, and similarly for all edges
$\{v^{\star},w\}_{w\in V}$.  

Assume $\lambdamc\in\Lambdamc_G$, and consider an edge $(u,v^{\star})\in
E_G$. Routing a message from $u$ to $v^{\star}$ in $G$ is implemented as
follows. Take the message at $u$ and split it into $n$ (distinct) parts
of equal length. Each part is to be sent to one of the $n$ nodes in $V$. In
other words, one part is kept at $u$, the other $n-1$ parts are sent
over the wireless network. This procedure is followed for every message
at every node $u\in V$. Note that the resulting traffic requirement is
unicast, and denote it by $\tlambdauc$. This unicast traffic matrix
$\tlambdauc$ is uniform, in the sense that each node $u\in V$ has
traffic for every other node $w\in V$ at the same rate, i.e., 
$\tlambdauc_{u,w}$ depends only on $u$ but is constant as a function of
$w$. Moreover, since $\lambdamc\in\Lambdamc_G$,
\begin{equation*}
    \sum_{w\neq u}\tlambdauc_{u,w} 
    \leq \sum_{\substack{W\subset V: \\ W\setminus\{u\}\neq\emptyset}}
    \lambdamc_{u,W}
    \leq 1
\end{equation*}
for every $u\in V$, and where we have the first inequality (instead of
equality) because one part of every message is kept at the source node
$u$.  Together, this implies that
\begin{equation*}
    \tlambdauc_{u,w} \leq 1/(n-1)
\end{equation*}
for all $u\neq w$, and we can assume without loss of generality that
we have equality for every $u\neq w$. This traffic pattern can be expressed
as a convex combination of $n-1$ permutation matrices, each of which can
be implemented at a rate of at least $2^{-\alpha/2}$ by using ergodic
interference alignment \cite{nazer09} as in the proof of Theorem
\ref{thm:unicast}. Hence, accounting for the factor $2$ loss due to time
sharing, all edges $\{(u,v^{\star})\}_{u\in V}$ in $G$ can be
implemented simultaneously with a loss of at most a factor
$2^{-1-\alpha/2}$ in the wireless network. 

Consider now an edge $(v^{\star},w)\in E_G$. Recall that all messages
originate at $V\subset V_G$, and hence to arrive at $v^{\star}$ in $G$
the message is distributed uniformly over the entire wireless network
(as described in the previous paragraph). Routing a message from
$v^{\star}$ to $w$ in $G$ can thus be implemented in the wireless
network by transmitting all the message parts from nodes $u\neq w$ to
$w$. We transmit this traffic as unicast traffic by duplicating all
messages that are to be sent to more than one destination node. Denote
again by $\tlambdauc$ the resulting unicast traffic matrix. Since the
messages are distributed uniformly, $\tlambdauc_{u,w}$ depends only on
$w$ but is constant as a function of $u$. Moreover, since
$\lambdamc\in\Lambdamc_G$, 
\begin{equation*}
    \sum_{u\neq w} \tlambdauc_{u,w} 
    \leq \sum_{u\neq w}\sum_{\substack{W\subset V: \\ w\in W}} \lambdamc_{u,W}
    \leq 1
\end{equation*}
for every $w\in V$.  Together, this implies that
\begin{equation*}
    \tlambdauc_{u,w} \leq 1/(n-1)
\end{equation*}
for all $u\neq w$, and we can assume again that we have equality for all
$u\neq w$. Expressing the resulting uniform traffic pattern as a convex
combination of permutation matrices and using again ergodic interference
alignment as in the previous paragraph shows that all edges
$\{(v^{\star},w)\}_{w\in V}$ in $G$ can be implemented simultaneously
with a loss of at most a factor $2^{-1-\alpha/2}$ in the wireless
network. 

Together this shows that if $\lambdamc\in\Lambdamc_G$ then
\begin{equation*}
    2^{-1-\alpha/2}\lambdamc\in\Lambdamc, 
\end{equation*}
and thus
\begin{equation}
    \label{eq:multicast4}
    \Lambdamc_G\subset 2^{1+\alpha/2}\Lambdamc.
\end{equation}
Combining \eqref{eq:multicast3} and \eqref{eq:multicast4} shows that
\begin{equation*}
    2^{-1-\alpha/2}\hLambdamc
    \subset 2^{-1-\alpha/2}\Lambdamc_G
    \subset\Lambdamc,
\end{equation*}
completing the proof of the inner bound. \qed

\section{Discussion}
\label{sec:discussion}

Here we discuss several aspects of the proposed communication schemes.
The dependence of the results on the network area $\card{A}$ is
discussed in Section~\ref{sec:discussion_area}. Implementation issues
are considered in Section~\ref{sec:discussion_delay}. Extensions to
Rayleigh fading (as opposed to phase fading) are discussed in
Section~\ref{sec:discussion_rayleigh}.

\subsection{Dependence on Network Area}
\label{sec:discussion_area}

Throughout this paper, we have assumed a unit network area, i.e.,
$\card{A}=1$. The results presented generalize to networks of area
\begin{equation*}
    \card{A}=\card{A(n)}\defeq a(n)
\end{equation*}
for general $a(n)$ depending on the number $n$ of nodes in the network.
Define the minimum distance between nodes to be
$r_{\min}(n)n^{-1/2}a^{1/2}(n)$; as before, we assume that $r_{\min}(n)$
decays at most polynomially in $n$.  Then Theorem~\ref{thm:unicast}
takes the form
\begin{equation*}
    \frac{1}{2}\log\big(1+ 2^{1-\alpha/2}a^{-\alpha/2}(n)\big)\hLambdauc(n) 
    \subset \Lambdauc(n) 
    \subset \log\big(1+n^{2+\alpha/2}r_{\min}^{-\alpha}(n)a^{-\alpha/2}(n)\big)
    \hLambdauc(n),
\end{equation*}
and Theorem~\ref{thm:multicast}
\begin{equation*}
    \frac{1}{4}\log\big(1+ 2^{1-\alpha/2}a^{-\alpha/2}(n)\big)\hLambdamc(n) 
    \subset \Lambdamc(n) 
    \subset \log\big(1+n^{2+\alpha/2}r_{\min}^{-\alpha}(n)a^{-\alpha/2}(n)\big)
    \hLambdamc(n).
\end{equation*}
Comparing the lower and upper bound in these two expressions, we see
that they provide the correct scaling of the unicast and multicast
capacity regions of the wireless network only if $a(n)=n^{o(1)}$, i.e.,
only if the region $A(n)$ grows slower than $n^{\beta}$ for any $\beta >
0$. This is not surprising, since when $a(n)$ grows on the order of
$n^{\beta}$ for $\beta > 0$, the network is no longer solely
interference limited, but rather also power limited. Under these
conditions, interference alignment is not the appropriate communication
strategy and some form of hierarchical cooperation
\cite{ozgur07b,niesen09,niesen08,ozgur08} or other form of cooperative
communication will likely be necessary (at least in the low $\alpha$
regime).

\subsection{Implementing Interference Alignment}
\label{sec:discussion_delay}

While the ergodic interference scheme recalled in Section
\ref{sec:schemes_alignment} is conceptually simple,
it suffers from very long coding delays for larger networks. Indeed, it
is easily seen that the coding delay of the scheme grows at least like
$\Omega(\exp(n^2))$. To be implemented, coding schemes whose delay
scales better with respect to the network size need to be used. Devising
such coding schemes guaranteeing the same rates as ergodic interference
alignment but with shorter delays would hence be of interest.

Similarly, the assumption of availability of full CSI at all nodes in
the network is quite strong. Relaxing this assumption would be of
interest. Some progress in this direction has been made in
\cite{gomadam08}, in which a distributed algorithm for interference
alignment using only local CSI is proposed. However, while this
algorithm is observed to yield good results in some scenarios, no
performance guarantee is given for general systems.

\subsection{Rayleigh Fading}
\label{sec:discussion_rayleigh}

Throughout this paper, we have assumed a simple phase-fading model
described by \eqref{eq:phase}. In this section, we discuss how the
results presented for this model can be adapted to the case of Rayleigh
fading. We will assume that the channel gains $\{h_{u,v}[t]\}$ are
independent (but not identically distributed) as a function of $u,v$ and
vary in a stationary ergodic manner in $t$. Each $h_{u,v}[t]$ is assumed
to be circularly-symmetric complex Gaussian with mean zero and variance
$r_{u,v}^{-\alpha}$. The realizations $\{h_{u,v}[t]\}_{u,v}$ are assumed
to be known at time $t$ throughout the network, i.e., we assume again
full CSI is available at all nodes.

Denote by $\Lambdamc(n)\subset\Rp^{n\times 2^n}$ the multicast capacity
region, and define $\hLambdamc(n)$ as in the phase-fading case [see
\eqref{eq:lambdamc}]. The next theorem approximates the multicast
capacity region under Rayleigh fading.
\begin{theorem}
    \label{thm:rayleigh}
    There exists $n_0$ such that for all $\alpha\geq 2$, $n\geq n_0$,
    and node placement $V(n)$ with minimum node separation
    $r_{\min}(n)n^{-1/2}$,
    \begin{equation*}
        \tfrac{1}{16}\big(\log\log(n) -\alpha/2-\log\log(e)\big)\hLambdamc(n) 
        \subset \Lambdamc(n) 
        \subset \log\big(4n^{2+\alpha/2}r_{\min}^{-\alpha}(n)\big)\hLambdamc(n).
    \end{equation*}
\end{theorem}

Comparing Theorem~\ref{thm:rayleigh} for Rayleigh fading with the
corresponding result Theorem~\ref{thm:multicast} for phase fading, we
see that the inner bound is enlarged by a factor of
$\Theta(\log\log(n))$. This is the gain due to opportunistic
communication enabled by the random amplitudes of the channel gains and
the availability of full CSI. Achievability is based on opportunistic
interference-alignment. Note that, since unicast traffic is a special
case of multicast traffic, Theorem~\ref{thm:rayleigh} also applies to
$\Lambdauc(n)$. 

\begin{IEEEproof}
    We first prove the outer bound. We assume throughout that $n\geq 9$.
    Following the same steps as in the proof of
    Theorem~\ref{thm:multicast}, it suffices to upper bound the MIMO
    capacities $C(\{w\}^c,\{w\})$ and $C(\{u\},\{u\}^c)$.  Relaxing
    again the individual power constraints to a sum-power constraint of
    $n-1$, and increasing the channel gains by multiplying each
    $h_{u,v}$ by 
    \begin{equation*}
        \frac{r_{\min}^{-\alpha/2}n^{\alpha/4}}{r_{u,v}^{-\alpha/2}}
        \geq 1,
    \end{equation*}
    we obtain
    \begin{equation*}
        C(\{w\}^c, \{w\}) 
        \leq \max\E\Big(\log\big(1+P(g)n^{\alpha/2} r_{\min}^{-\alpha} g\big)\Big),
    \end{equation*}
    where
    \begin{equation*}
        g \defeq \sum_{u\neq w}
        \abs{r_{u,v}^{\alpha/2}h_{u,w}}^2,
    \end{equation*}
    and where the maximization is over all power assignments $P(g)$
    such that
    \begin{equation*}
        \E \big(P(g)\big) \leq n-1.
    \end{equation*}
    By \cite{goldsmith97}, this maximization problem is solved by water
    filling. The optimal power allocation is
    \begin{equation*}
        P^\star(g)
        = \Big(\frac{1}{g_0}-\frac{1}{n^{\alpha/2}r_{\min}^{-\alpha}g}\Big)^+,
    \end{equation*}
    with $g_0$ chosen such that
    \begin{equation*}
        \E\big(P^\star(g)\big) = n-1.
    \end{equation*}
    
    Noting that
    \begin{equation*}
        P^\star(g) \leq \frac{1}{g_0},
    \end{equation*}
    we can upper bound
    \begin{align}
        \label{eq:rayleigh1}
        C(\{w\}^c, \{w\}) 
        & \leq \E\log\big(1+P^\star(g)n^{\alpha/2}r_{\min}^{-\alpha}g\big) \nonumber\\
        & \leq \E\log\big(1+n^{\alpha/2}r_{\min}^{-\alpha}g/g_0\big) \nonumber\\
        & \leq \log\big(1+n^{\alpha/2}r_{\min}^{-\alpha}\E(g)/g_0\big) \nonumber\\
        & \leq \log\big(1+n^{1+\alpha/2} r_{\min}^{-\alpha}/g_0\big),
    \end{align}
    where we have used Jensen's inequality. 
    
    It remains to find a lower bound on $g_0$. From the power
    constraint,
    \begin{align}
        \label{eq:rayleigh2}
        n-1
        & = \E\Big(\frac{1}{g_0} -\frac{1}{n^{\alpha/2}r_{\min}^{-\alpha}g}\Big)^+ \nonumber\\
        & = \int_{\gamma=n^{-\alpha/2}r_{\min}^{\alpha}g_0}^{\infty}f_g(\gamma)
        \Big(\frac{1}{g_0}-\frac{1}{n^{\alpha/2}r_{\min}^{-\alpha}\gamma}\Big)
        d\gamma \nonumber\\
        & \geq \frac{1}{2g_0}\Pp(g\geq 2n^{-\alpha/2}r_{\min}^{\alpha}g_0) \nonumber\\
        & \geq \frac{1}{2g_0}\Pp(g\geq 2g_0),
    \end{align}
    where we have used that $r_{\min} \leq 3$ by \eqref{eq:rmin} and
    that $n\geq 9$ by assumption.  The random variable $g$ is the sum of
    $n-1$ i.i.d. exponential random variables with mean one.  Hence $g$
    follows an Erlang distribution with density
    \begin{equation*}
        f_g(\gamma) = \frac{\gamma^{n-2}\exp(-\gamma)}{(n-2)!}
    \end{equation*}
    and
    \begin{equation*}
        \Pp(g \geq \gamma) =
        \exp(-\gamma) \sum_{i=0}^{n-2}\frac{\gamma^i}{i!},
    \end{equation*}
    both for $\gamma \geq 0$. From this, 
    \begin{align*}
        \frac{1}{2g_0}\Pp(g\geq 2g_0)
        & = \frac{\exp(-2g_0)}{2g_0}\sum_{i=0}^{n-2}\frac{(2g_0)^i}{i!} \nonumber\\
        & \geq \frac{\exp(-2g_0)}{2g_0}.
    \end{align*}
    Combined with \eqref{eq:rayleigh2}, we obtain
    \begin{equation}
        \label{eq:rayleigh3a}
        n-1 \geq \frac{1}{2g_0}\exp(-2g_0).
    \end{equation}
    Assume $g_0 < 1/4(n-1)$; then 
    \begin{align*}
        \frac{1}{2g_0}\exp(-2g_0)
        & > 2(n-1)\exp(-1/2(n-1)) \\
        & \geq 2(n-1)\exp(-1/2) \\
        & \geq (n-1),
    \end{align*}
    contradicting \eqref{eq:rayleigh3a}. This shows that
    \begin{equation}
        \label{eq:rayleigh4}
        g_0 \geq 1/4(n-1).
    \end{equation}

    Combining \eqref{eq:rayleigh1} and \eqref{eq:rayleigh4},
    \begin{align*}
        C(\{w\}^c, \{w\}) 
        & \leq \log\big(1+4(n-1)n^{1+\alpha/2} r_{\min}^{-\alpha}\big) \\
        & \leq \log\big(4n^{2+\alpha/2} r_{\min}^{-\alpha}\big)
    \end{align*}
    for every $w\in V$, and for $n\geq 9$. Similarly
    \begin{equation*}
        C(\{u\}, \{u\}^c) 
        \leq \log\big(4n^{2+\alpha/2} r_{\min}^{-\alpha}\big)
    \end{equation*}
    for every $u\in V$. This proves the outer bound on $\Lambdamc$.

    We continue with the proof of the inner bound. From the construction
    in the proof of Theorem~\ref{thm:multicast}, it suffices to analyze
    communication according to the unicast traffic matrix
    $\lambdauc_{u,w}=\rho(n)$ for all $u\neq w$, for some $\rho(n)$
    depending on $n$ but not on $u,w$. If this $\lambdauc$ is achievable
    for some $\rho(n)$, then 
    \begin{equation}
        \label{eq:rayleigh5}
        \tfrac{n}{2}\rho(n) \hLambdamc \subset \Lambdamc.
    \end{equation}

    Construct an undirected graph $\tilde{G}[t] = (V_{\tilde{G}}[t],
    E_{\tilde{G}}[t])$ as follows. The vertex set $V_{\tilde{G}}[t]$ is equal
    to the collection of nodes $V$ in the wireless network for every
    $t\in\N$. The edge $(u,v)$ is in $E_{\tilde{G}}[t]$ if
    \begin{equation*}
        \max\big\{\abs{h_{u,v}[t]}^2,\abs{h_{v,u}[t]}^2\big\} 
        \geq \ln(1/p(n)) r_{u,v}^{-\alpha},
    \end{equation*}
    with 
    \begin{equation*}
        p(n) \defeq 1/\sqrt{n}.
    \end{equation*}
    Note that $\abs{r_{u,v}^{\alpha/2}h_{u,v}[t]}^2$ is exponentially
    distributed with unit mean, and hence
    \begin{equation*}
        \Pp\big(\abs{h_{u,v}[t]}^2
        \geq \ln(1/p(n)) r_{u,v}^{-\alpha}\big)
        = p(n)
    \end{equation*}
    for every $u,v\in V$ with $u\neq v$. Thus $\tilde{G}[t]$ is a random
    graph with $n$ vertices and each edge present i.i.d. with
    probability $p(n)$. 
    
    The choice of $p(n)$ guarantees by \cite[Theorem 7.14]{bollobas01}
    that, with probability $1-o(1)$ as $n\to\infty$, the graph
    $\tilde{G}[t]$ has a matching covering at least $n-1$ vertices,
    i.e., there is at least one way to pair adjacent nodes in
    $\tilde{G}[t]$ such that that all (except for possibly one node if
    $n$ is odd) nodes in $V$ are member of exactly one
    pair.\footnote{The precise threshold for the appearance of such a
    matching is, in fact, for $p(n)$ larger than
    $(\log(n)+\omega(1))/n$. However, the weaker choice of $p(n) =
    1/\sqrt{n}$ adopted here is sufficient for our purposes.} Choose
    $n_0$ such that this probability is at least $1/2$ for $n\geq n_0$.
    Whenever no such pairing exists, we do not communicate during that
    time slot; this yields a factor $2$ loss in rate. Assume in the
    following that at least one such pairing exists.  Pick one of the
    (possibly many) pairings at random. By construction of
    $E_{\tilde{G}}$, for every such pair $(u,v)$ either
    $\abs{h_{u,v}[t]}^2$ or $\abs{h_{v,u}[t]}^2$ is larger than 
    \begin{equation*}
        \ln(n)r_{u,v}^{-\alpha}/2
        \geq 2^{-1-\alpha/2}\ln(n).
    \end{equation*}
    For each pair $(u,v)$ choose $u$ as the source for $v$ if the
    magnitude of the channel gain from $u$ to $v$ is larger than from
    $v$ to $u$, and $v$ as a source for $u$ otherwise.
    During time $t$ we transmit according to this
    source-destination pairing at uniform rate $\tilde{\rho}[t]$.

    Consider now all times $t$ that have resulted in the same
    source-destination pairing. Note that the construction of
    $\tilde{G}[t]$, and hence also the construction of the
    source-destination pairing, depends only on the magnitudes of the
    channel gains $h_{u,v}[t]$. Hence, conditioned on a particular
    realization of $\tilde{G}[t]$, the phases of the channel gains are
    still independently and uniformly distributed over $[0,2\pi)$ for
    every $u,v\in V$. The fading, conditioned on the source-destination
    pairing resulting from $\tilde{G}[t]$, is therefore still
    circularly-symmetric, and we can hence apply ergodic interference
    alignment as in Theorem~\ref{thm:alignment} to communicate at
    uniform rate
    \begin{align*}
        \tilde{\rho}[t] 
        & \geq \frac{1}{2}\log\big(1+2^{-\alpha/2}\ln(n)\big) \\
        & \geq \frac{1}{2}\big(-\alpha/2-\log\log(e)+\log\log(n)\big).
    \end{align*}

    During each time $t$, at least $(n-1)/2$ of the source nodes are
    transmitting at rate at least $\tilde{\rho}[t]$. By the random choice of
    source-destination pairing, over a long enough time period all node pairs
    $(u,w)$ communicate the same fraction of time. Accounting for the
    half of time slots during which the graph $\tilde{G}[t]$ has no
    valid pairing, this procedure
    achieves a rate between each of the $n(n-1)$ pairs $(u,w)$ with
    $u\neq w$ of at least
    \begin{align*}
        \rho(n)  
        & \geq \frac{n-1}{4n(n-1)} \Big(\frac{1}{2}\big(-\alpha/2-\log\log(e)
        +\log\log(n)\big)\Big) \\
        & = \frac{1}{8n}\big(-\alpha/2-\log\log(e)
        +\log\log(n)\big).
    \end{align*}
    By \eqref{eq:rayleigh5}, this implies that for, $n\geq n_0$,
    \begin{equation*}
        \frac{1}{16}\big(-\alpha/2-\log\log(e)
        +\log\log(n)\big) \hLambdamc(n) 
        \subset \Lambdamc(n),
    \end{equation*}
    concluding the proof of achievability. 
\end{IEEEproof}

\section{Conclusions}
\label{sec:conclusions}

We presented inner and outer bounds on the $n\times n$-dimensional
unicast capacity region $\Lambdauc(n)$ and the $n\times 2^n$-dimensional
multicast capacity region $\Lambdamc(n)$ of a dense wireless network
with $n$ nodes placed arbitrarily on a unit square. These bounds are
tight up to a factor $O(\log(n))$ (with a pre-constant that is rather
small), and hence they yield fairly tight scaling laws for achievable
rates under any unicast or multicast traffic pattern and any node
placement.

\section{Acknowledgments}

I would like to thank B. Nazer for helpful discussions and the anonymous
reviewers for their comments.

\end{document}